\documentclass{article}

\PassOptionsToPackage{numbers, compress}{natbib}

 \usepackage[preprint]{neurips_2020_tda}

\usepackage[utf8]{inputenc}
\usepackage[T1]{fontenc}
\usepackage{amsfonts}
\usepackage{amsmath}
\usepackage[colorlinks=true]{hyperref}

\usepackage{graphicx}
\usepackage{subcaption}

\title{
    Interpretable Phase Detection and Classification\\ with Persistent Homology
}

\author{
    Alex Cole \\
    Gravitation Astroparticle Physics Amsterdam (GRAPPA)\\
    Institute for Theoretical Physics Amsterdam\\
    University of Amsterdam, the Netherlands \\
    \texttt{a.e.cole@uva.nl} \\
    \And
    Gregory J.\ Loges \\
    Department of Physics \\
    University of Wisconsin-Madison, USA \\
    \texttt{gloges@wisc.edu} \\
    \And
    Gary Shiu \\
    Department of Physics \\
    University of Wisconsin-Madison, USA \\
    \texttt{shiu@physics.wisc.edu} \\
}

\begin{document}

\maketitle

\begin{abstract}
    We apply persistent homology to the task of discovering and characterizing phase transitions, using lattice spin models from statistical physics for working examples.
    Persistence images provide a useful representation of the homological data for conducting statistical tasks.
    To identify the phase transitions, a simple logistic regression on these images is sufficient for the models we consider, and interpretable order parameters are then read from the weights of the regression.
    Magnetization, frustration and vortex-antivortex structure are identified as relevant features for characterizing phase transitions.
\end{abstract}

\section{Introduction}

Phase transitions in physics are many and varied. They can be associated with the breaking of some symmetry or indicate that the natural degrees of freedom have changed, and order parameters characterizing the phase transition can give deep insight into the physical system and its underlying dynamics.
The task of understanding the phase structure of a physical system may be easily phrased in terms of machine learning.
Namely, ``how many phases are there'' is a question addressed by \emph{unsupervised learning}, while the question ``how are different phases distinguished'' is an exercise in \emph{supervised learning}.

The study of phase transitions in lattice spin models using machine learning has been quite successful (see~\cite{Wang_2016,van_Nieuwenburg_2017,Wetzel_2017,Hu_2017,Woloshyn:2019oww,Giannetti_2019,Carrasquilla_2017,Ch_ng_2017,Huembeli2017:1710.08382v2,Tanaka2016:1609.09087v2,Carrasquilla_2020}, among others).
Clustering algorithms, support vector machines and (deep) neural networks have all proven useful in understanding such phenomena.
In this paper we demonstrate how persistent homology can be used as a tool for detecting and characterizing phase transition in a way that is both amenable to statistical analysis and lends itself to a straightforward physical interpretation (see~\cite{donato2016persistent,2020arXiv200102616S,tran2020topological,olsthoorn2020finding} for related work). This paper is a condensed version of~\cite{Cole:2020hjx}.

\section{Methods}

Given a spin configuration for a particular lattice system there are different ways that one may interpret the spins in a way appropriate for analysis with persistent homology.
For models with discrete spins we take as a point cloud the physical locations of those spins aligned with the total magnetization (no matter how small).
An $\alpha$-filtration~\cite{10.1145/174462.156635} is then used to extract the topological features of the spin configuration.
For models with continuous spins taking values in $S^1$, we think of a spin configuration as a map $f:\Lambda\to (-\pi,\pi]$ from the lattice of spin sites to a finite interval and use a sublevel filtration of cubical complexes.

For the statistical tasks of identifying phase transitions, order parameters and critical phenomena we employ \emph{persistence images}~\cite{adams2015persistence} as vectorized representations of the persistence data.
Persistence images are formed by smoothing and binning the persistence data in the birth-persistence plane, using a weight factor which goes to zero for vanishing persistence, e.g.\ $\log{(1+p)}$, so as to emphasize those longer-lived features and be robust against statistical noise. For more on the stability properties of persistence images, see~\cite{adams2015persistence}.

\section{Results: phase classification \& critical phenomena}

We consider three simple two-dimensional lattice spin models: the Ising and square-ice models have discrete, $\pm$, spins and the XY model has continuously varying spins.
Standard Monte-Carlo sampling techniques are used to generate 1000 sample configurations at a number of temperatures.

For each model considered, classification into two phases is performed using only the persistence images.
Because persistence images are information-rich, we are able to use perhaps the simplest classification scheme, logistic regression, to great effect.
Other classification techniques can be used to refine the analysis, but even this linear method is successful when used with persistence images.
A subset of samples with extreme temperatures is used to train a logistic regression with $\ell_2$-regularization and then the accuracy of the regression is evaluated using the known temperatures of all samples.
The accuracy is largely insensitive to the choice of extreme training temperatures.
We normalize our persistence images using the $\ell_1$-norm, so they may be interpreted as probability densities for finding cycles with particular births/deaths for a given system.

\begin{figure}[t]
	\centering
	\includegraphics[width=\textwidth]{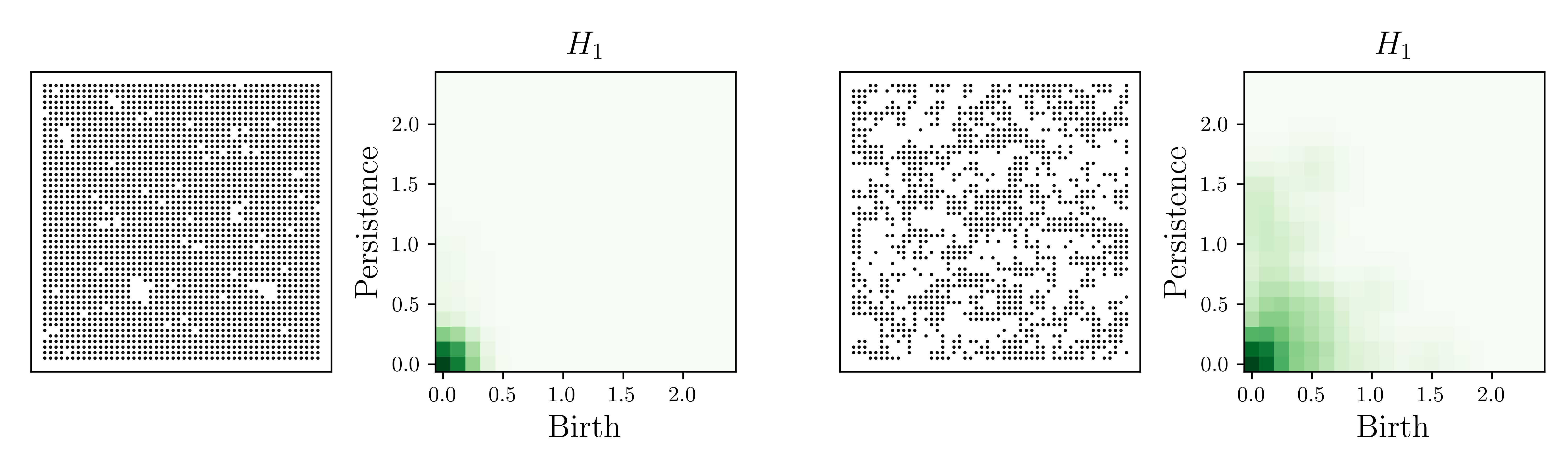}
	\vspace{-15pt}
	
	\includegraphics[width=\textwidth]{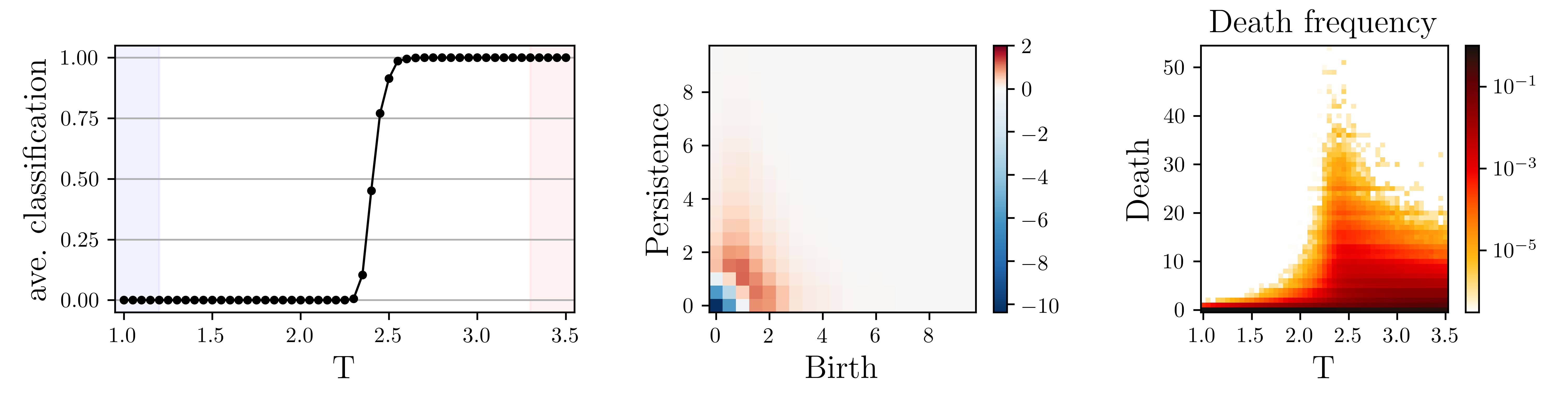}
	\caption{(Top) Two Ising spin configurations ($T=1.9$ and $T=3.5$) and their persistence images. (Bottom left) Average classification, with training data in the highlighted regions only. (Bottom center) Trained coefficients: blue (red) regions are characteristic of low (high) temperatures. (Bottom right) Distribution of 1-cycle deaths.}
	\label{fig:Ising}
\end{figure}

\paragraph{Ising model}
The Ising model on a two-dimensional square lattice is very well understood, largely in part to Onsager's exact solution~\cite{PhysRev.65.117}.
Spins $s_i\in\{{-1},1\}$ live at the vertices of the lattice with ferromagnetic interactions governed by the local Hamiltonian $H_\text{Is} = -\sum_{\langle i,j\rangle} s_is_j$, where the sum is over nearest-neighbor pairs.
In the thermodynamic limit there is a second-order phase transition at $T_\text{Is}=\frac{2}{\log{(1+\sqrt{2})}}\approx2.27$.
At low temperatures there is spontaneous magnetization, while there is a disordered phase at high temperatures.
While this model is well-understood, it provides a good first application of our method.

In Figure~\ref{fig:Ising} we see that the classification using logistic regression extrapolates very well to the intermediate temperatures and gives an estimate of $T\approx2.37$ for the critical temperature.
The coefficients of the trained logistic regression show that the low-temperature configurations are identified by their having many small, short-lived cycles.
These may be understood as arising both from $2\times2$ blocks of aligned spins (which lead to very short-lived 1-cycles) as well as 1-cycles wrapping small groups of isolated spins which are flipped relative to the large domains of aligned spins: the latter become more and more important as the temperature is increased.
In the high-temperature phase, spins are oriented randomly, leading to a more uniform distribution of 1-cycle sizes.
Using persistent homology we are able to easily identify the magnetization as the order parameter, as is well known.

We are also able to see aspects of scale-invariance appearing at criticality by looking at statistics derived from the persistence diagram.
We may compute the 1-cycle death probability density, $\mathrm{D}_T(d)$, at each temperature, which quantifies the distribution of feature sizes in the spins.
As shown in Figure~\ref{fig:Ising}, we find that deaths are exponentially distributed with a long tail forming at criticality, indicative of a diverging correlation length and the emergence of power-law behavior.
In fact, we may fit each $\mathrm{D}_T(d)$ to a function of the form $\mathrm{D}_T(d) = A\,d^{-\mu}e^{-d/a_\text{death}}$ and extract the critical exponents $\mu$ and the degree of divergence, $\nu_\text{death}$, for the ``correlation area'' $a_\text{death}$, defined by $a_\text{death}\sim|T-T_\text{c}|^{-\nu_\text{death}}$.
These give quantitative ways to characterize the Ising critical point.
One finds that $\mu\approx2$ and $\nu_\text{death}\approx1$, in agreement with expectations from the distribution of cluster sizes at criticality, which follows a power-law $(\text{cluster size})^{-2.032}$~\cite{1983JPhA.16.1721B,Toral:1987ba}, and the degree of divergence of the spin-spin correlation length, $\xi\sim|T-T_\text{c}|^{-1}$. It would be interesting to develop this connection to known critical exponents further.

\begin{figure}[t]
	\centering
	\includegraphics[width=\textwidth]{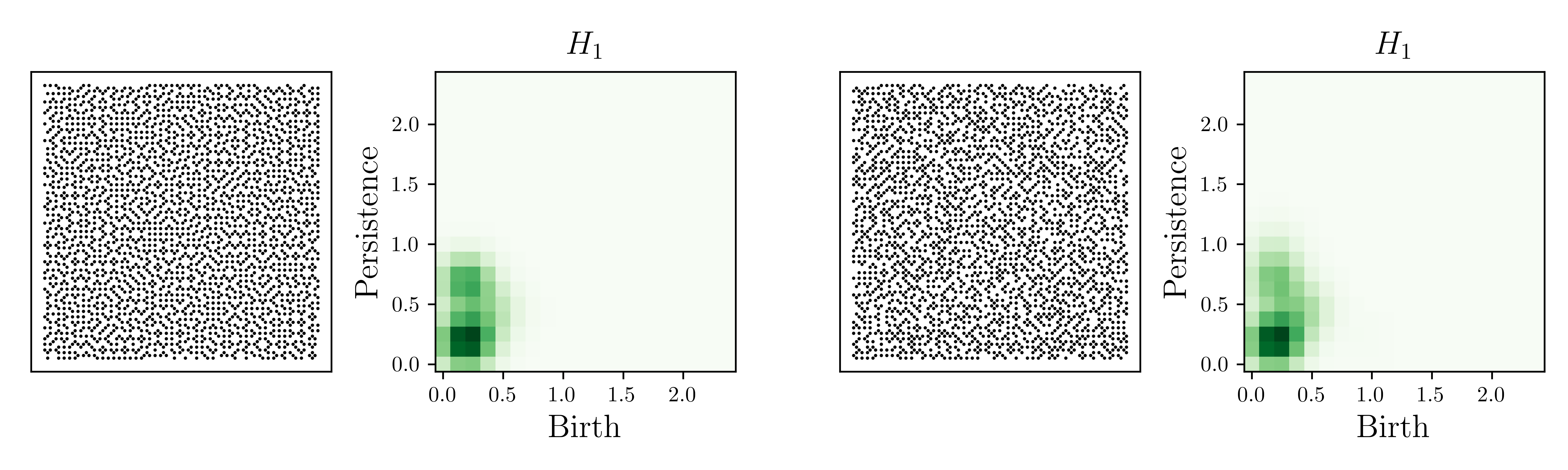}
	\vspace{-20pt}
	
	\includegraphics[width=\textwidth]{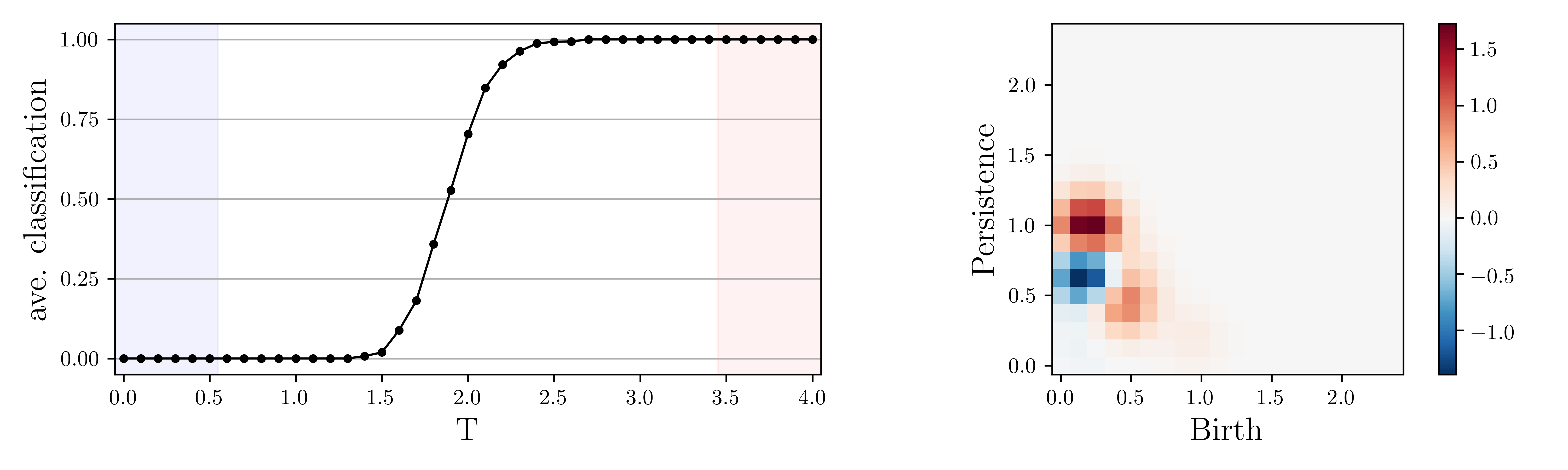}
	\caption{(Top) Two square-ice spin configurations ($T=0.5$ and $T=4.0$) and their persistence images. (Bottom left) Average classification, with training data in the highlighed regions only. (Bottom right) Trained coefficients: blue (red) regions are characteristic of low (high) temperatures.}
	\label{fig:SquareIce}
\end{figure}

\paragraph{Square-ice model}
The square-ice model places spins, $s_i\in\{{-1},1\}$, on the edges rather than vertices of a square lattice and is governed by the local Hamiltonian $H_\text{SI} = \sum_{v\in\Lambda}(\sum_{i:v}s_i)^2$, where $i:v$ denotes those spins on edges adjacent to the vertex $v$.
In contrast to the Ising model there is no spontaneous magnetization at low temperatures.
Rather, the ground state is highly degenerate: any configuration with exactly two up and two down spins adjacent to every vertex has zero energy.

Again training a logistic regression only on those persistence images with extreme temperatures (see Figure~\ref{fig:SquareIce}), we find an estimate of $T\approx 1.9$ for the critical temperature.
From the logistic regression coefficients we see that as the temperature increases there is a tendency for 1-cycles to be born later or to be longer-lived.
Both are indicative of a changing local structure in the spin configurations.
In the low-temperature phase, it is energetically beneficial for neighboring vertices to coordinate, resulting in a regular patterns of alternating up and down spins.
This regularity forces 1-cycles to live at smaller scales than in the higher-temperature phase.

\begin{figure}[t]
	\centering
	\includegraphics[width=\textwidth]{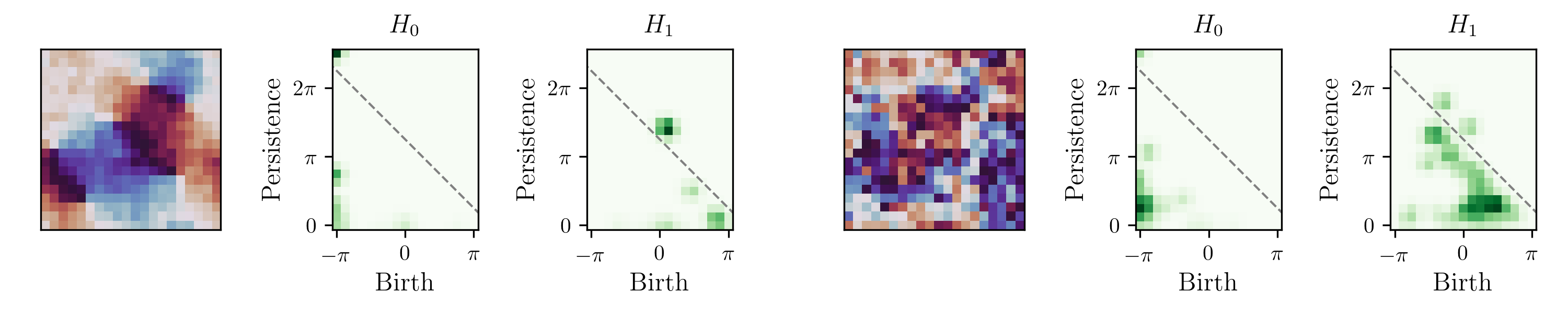}
	\vspace{-15pt}
	
	\includegraphics[width=\textwidth]{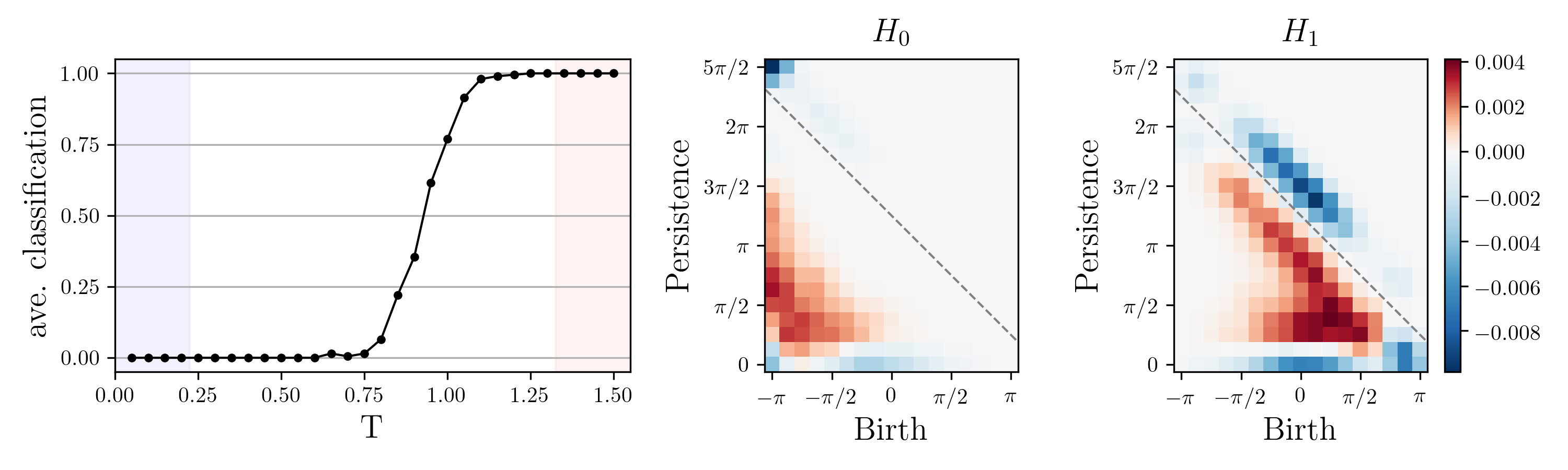}
	\caption{(Top) Two XY spin configurations ($T=0.1$ and $T=1.4$) and their persistence images. (Bottom left) Average classification, with training data in the highlighted regions only. (Bottom right) Trained coefficients: blue (red) regions are characteristic of low (high) temperatures.}
	\label{fig:XY}
\end{figure}

\paragraph{XY model}
The XY model is a continuous-spin generalization of the Ising model.
At each site of the square lattice spins take values in $S^1$ and are governed by $H_\text{XY} = -\sum_{\langle i,j\rangle}\cos{(\theta_i-\theta_j)}$.
There is a well-known Kosterlitz-Thouless phase transition at $T_\text{XY}\approx 0.892$ (see~\cite{PhysRevB.37.5986,Hasenbusch:2005xm}, among others).
This is an infinite-order phase transition where at low temperatures there are bound vortex-antivortex pairs while at high temperatures free vortices proliferate and spins are randomly oriented.

The zeroth homology, in contrast to the $\alpha$-complexes used for discrete spins, is very rich for the cubical complexes and we include both $H_0$ and $H_1$ persistence data in the persistence images.
The infinite-persistence $p$-cycles corresponding to the torus on which the lattice lives are distinguished from those cycles with late death ($d\approx\pi$) by giving the former a death of $d=\frac{5\pi}{2}$ by hand.
Omitting these immortal ``torus cycles'' results in a comparable phase classification.

Performing a logistic regression on the concatenated $H_0$ and $H_1$ persistence images by training on configurations with temperatures far away from the anticipated transition leads to the classification of Figure~\ref{fig:XY} and estimate of $T_\text{XY}\approx0.9$.
We see that the low-temperature phase is characterized by $p$-cycles on the ``boundary'' of the persistence images.
This we can understand in the following way.
Vortex-antivortex pairs entail having nontrivial winding around the $S^1$ and so have spins of extreme angles ($\theta\approx\pm\pi$).
This produces 0-cycles with early birth, one of which survives forever: this explains the two blue corners on the left-side of the $H_0$ coefficients.
There are also nontrivial 1-cycles around spins close to $\theta\approx\pi$ that explain the strong blue region in the bottom right of the $H_1$ coefficients.

\section{Discussion}

Persistent homology is a useful quantitative tool in understanding the finer details of phase transitions and critical phenomena.
We have demonstrated its utility for lattice spin models, where it easily identifies magnetization in the Ising model, changing local structure in the square-ice model, and vortex-antivortex pairs in the XY model.
For the second-order phase transition of the Ising model we have shown that persistent homology captures the approach towards criticality via two critical exponents which we estimate. Statistical physics systems are an ideal context for building and understanding machine learning, in particular persistent homology, pipelines. The ``data manifold'' is described explicitly by a Boltzmann distribution $P(x)\sim e^{-\beta H(x)}$ (or, depending on what is held fixed, a similar probability distribution with some other free energy replacing $H$). Additionally, given that renormalization group techniques were originally derived in the context of statistical physics~\cite{RevModPhys.55.583}, these systems are naturally suited to explore how machine learning pipelines utilize multiscale information to form internal representations of data sets.

\begin{ack}
    We thank Jeff Schmidt for useful discussions. We also thank the participants of the ``Theoretical Physics for Machine Learning'' and ``Physics $\cap$ ML'' workshops at the Aspen Center for Physics and Microsoft Research, respectively, where partial results of this work were presented in early 2019, for discussion.
    G.J.L.\ and G.S.\ are supported in part by the DOE grant DE-SC0017647 and the Kellett Award of the University of Wisconsin.
    The code and data used in our analysis are provided in the following GitHub repository: \texttt{gloges/TDA-Spin-Models}.
\end{ack}

\bibliographystyle{utphys}
\bibliography{TDA_spins}

\end{document}